\documentclass[12pt,preprint]{aastex}

\usepackage{natbib}
\def\degree{\ifmmode {^\circ}\else {$^\circ$}\fi}
\def\rstar{\ifmmode {\, R_{\star}}\else $R_{\star}$\fi}
\def\msol{\ifmmode {\, M_{\odot}}\else $M_{\odot}$\fi}
\def\rsol{\ifmmode {\, R_{\odot}}\else $R_{\odot}$\fi}
\def\lsol{\ifmmode {\, L_{\odot}}\else $L_{\odot}$\fi}
\def\msolyr{\ifmmode {\, M_{\odot}\,{\rm yr}^{-1}}\else $M_{\odot}\,{\rm yr}^{-1}$\fi}
\def\mdot{\ifmmode {\,\dot{M}}\else $\dot{M}$\fi}
\def\mdotyr{\ifmmode {\,\dot{M}\,yr^{-1}}\else $\dot{M}\,yr^{-1}$\fi}

\newcommand{\Teff}{\ifmmode{T_{\rm eff}}\else{$T_{\rm eff}$}}

\begin{document}

\title{V532 Oph is a New R Coronae Borealis Star}

\author{Geoffrey C. Clayton\altaffilmark{1}, D. Kilkenny\altaffilmark{2}, P. Wils\altaffilmark{3}, and D.L. Welch\altaffilmark{4}}

\altaffiltext{1}{Department of Physics \& Astronomy, Louisiana State
University, Baton Rouge, LA 70803; gclayton@fenway.phys.lsu.edu}

\altaffiltext{2}{Department of Physics, University of the Western Cape,
 Private Bag X17, Bellville 7535, South Africa; dkilkenny@uwc.ac.za}

\altaffiltext{3}{Vereniging Voor Sterrenkunde, Belgium; patrickwils@yahoo.com}

\altaffiltext{4}{Dept. of Physics \& Astronomy, McMaster University, Hamilton, Ontario,  L8S 4M1 Canada; welch@physics.mcmaster.ca}

\begin{abstract}
V532 Oph has been found to be a member of the rare, hydrogen-deficient R Coronae Borealis (RCB) stars from new photometric and spectroscopic data reported in this paper. 
The lightcurve of V532 Oph shows the sudden, deep, irregularly spaced declines characteristic of RCB stars. Its optical spectrum is typical of a warm (T$_{eff}\sim$ 7000 K) RCB star, showing weak or absent hydrogen lines, the C$_2$ Swan bands, and no evidence for $^{13}$C. In addition, the star shows small pulsations typical of an RCB star and an infrared excess due to circumstellar dust. It also appears to be significantly reddened by foreground dust. The distance to V532 Oph is estimated to be 5.5-8.7 kpc. 
These new data show that this star was misclassified as an eclipsing binary in the General Catalog of Variable Stars.
The new data presented here for V532 Oph
reveal the power of high-quality, high-cadence all-sky
photometric surveys, such as ASAS-3, to identify new RCB
candidates on the basis of lightcurve data alone, now that
they have been collecting data for durations sufficiently
long to reveal multiple declines.
Despite their small numbers, RCB stars may be of great importance in understanding the late stages of stellar evolution.  In particular, their measured isotopic abundances imply that many, if not most, RCB stars are produced by WD mergers, which may be the  low-mass counterparts of the more massive mergers thought to produce type Ia supernovae. Therefore, establishing the population of RCB stars in the Galaxy will help constrain the frequency of these WD mergers. 

\end{abstract}



\section{Introduction}

The R Coronae Borealis (RCB) stars are a rare class of extremely interesting transition objects which
have the potential to reveal critical details of the late stages of stellar evolution. 
These stars form a small group of carbon-rich supergiants which
are defined by extreme hydrogen deficiency and unusual variability. RCB stars undergo large declines of up to 8 mag due to the formation of carbon dust at irregular intervals. Two scenarios have been proposed for the origin of an RCB star: the Double Degenerate (DD) and the final helium-shell flash (FF) models \citep{1996ApJ...456..750I,2002MNRAS.333..121S}. The former involves the merger of a CO- and a He-WD \citep{1984ApJ...277..355W}. In the latter, a star evolving into a planetary nebula (PN) central star expands to supergiant size by a FF \citep{1977PASJ...29..331F,1979sss..meet..155R}.  \citet{2005ApJ...623L.141C,2007ApJ...662.1220C} found that some RCB stars have $^{18}$O/ $^{16}$O ratios that are orders of magnitude higher than those seen for any other known class of stars which favors the WD merger scenario while four RCB stars, including R CrB itself,  show enhanced Li abundances, which favors the FF scenario. 
The obvious conclusion is that there are (at least) two evolutionary channels leading to the RCB stars, perhaps with the DD being the dominant mechanism.


Establishing the population of RCB stars in the Galaxy will place strong constraints on the lifetimes of these stars as well as their birthrate which will, in turn, help to constrain the FF and DD scenarios. It has been predicted that there may be as many as ~3000 RCB stars in the Galaxy as a whole based on the numbers found in the Magellanic Clouds but only about 65 RCB stars have been discovered so far \citep{2001ApJ...554..298A,2004A&A...424..245T,2005AJ....130.2293Z}. The pace of discovery of RCB stars in the Galaxy has been accelerating with about 25 new RCB stars identified in the last 15 years \citep{1994AJ....108..247B,2002PASP..114..846C,2003PASP..115.1301H, 2005AJ....130.2293Z,2008A&A...481..673T}.
A more accurate knowledge of the birthrates of RCB stars will constrain the frequency of low-mass WD mergers that produce them in the DD scenario.
This will help us to establish the rates of more massive WD mergers that are thought to produce type Ia supernovae. 

In the General Catalog of Variable Stars, V532 Oph is classified as an Algol-type eclipsing binary \citep{2009yCat....102025S}. This identification was made by \citet{1942AnHar.109....1S}, who found that the brightness of this star (= SVS 457) was constant from 1890-1939 except for one large drop in brightness that occurred in 1928 June ($\sim$JD 2425413). At that time, V532 Oph dropped  from its maximum, m$_{pg}\sim$12.6 mag,  to $>$15.5 mag. No other significant variation in brightness was observed over 40 years. However, new ASAS-3 telescope photometry, covering the period 2001 to 2009, shows that V532 Oph is not an eclipsing binary system. Its lightcurve shows large declines in brightness reminiscent of an RCB star. In this paper, we present  and interpret newly acquired photometry and spectroscopy of V532 Oph which show that this star is indeed an RCB star.

\section{Observations and Data Reduction}

V532 Oph lies at $\alpha$(2000) 17$^h$ 32$^m$ 42\fs61, $\delta$(2000) 
-21\arcdeg~51\arcmin~40\farcs76 (UCAC2 23179321) \citep{2004AJ....127.3043Z}. Figure 1 shows a chart with V532 Oph identified. 
Figure 2a shows the $m_{pg}$ measurements presented in \citet{1942AnHar.109....1S}. 
 The 
uncertainty in the $m_{pg}$ measurements is on the order of $\pm$0.2~mag based on the scatter in the points.  
This photometry was measured on various plate/instrument/telescope combinations so is not necessarily very uniform across 1890-1939 timeframe. Most of the plate material was blue.
It is unknown whether the fading of V532 Oph seen in the very early data is real.
However, the decline in brightness seen $\sim$JD 2425362 (1928 April 25) is real. 
Figure 2b shows V-band data from the ASAS-3 telescope \citep{2002AcA....52..397P}. The scale is 14\arcsec per pixel.  A bright star, HD 15882 (V$\sim$8 mag), lies 1\arcmin~from V532 Oph (See Figure 1) so a small aperture must be used for the photometry. Therefore, we have plotted the ASAS-3 V-band photometry using the 2-pixel aperture. 
The uncertainties in the ASAS-3 points are typically $\sim$0.05 mag. The scatter in the photometry seems larger than this. A 5-point boxcar smooth has been applied to the data plotted in Figure 2b. 
Also plotted in Figure 2b are two V-band data points downloaded from the AAVSO International Database (Henden, A.A., 2008, private communication). 
BVRI photometry was obtained on JD 2454674 and 2454677. The two epochs are only three days apart and the star did not appear to vary significantly in that time. Both observations give V= 12.8, B-V=1.1, V-R=0.4, V-I=0.85.

New spectroscopic observations of V532 Oph were obtained 
on 2008 September 3-8, when it was about 0.8 mag below maximum while recovering from a decline. The epoch of these spectra is marked on the lightcurve in Figure 2b. 
The spectra were obtained with the
SIT/CCD spectrograph on the SAAO 1.9m telescope at Sutherland, South 
Africa using
grating 6 which has a 
resolution of $\sim$2.5 \AA,
and a useful range of about
3600-5400~\AA\ at the angle setting used.
The spectra have been extracted,
flat-field corrected, sky-subtracted and wavelength calibrated.
A flux calibration was done for the spectra using the spectrophotometric standard LTT7379 although due to the slit and variable seeing, the data are not photometric. 
Eight spectra were obtained of V532 Oph. These were combined by taking a median of all the spectra. This median spectrum is shown in Figure 3 along with spectra at similar resolution of two RCB stars, W Men and HV 12842. The spectra of W Men and HV 12842 were also obtained with the  SAAO 1.9m in 1997 November using the same setup described above.

\section{Discussion}
Other than what is shown in Figure 2, there is little other photometry of V532 Oph in the literature. 
The transformation from $m_{pg}$ to Johnson $V$ ($V-m_{pg}$ = 0.17 - 1.09$\times$($B-V$)) gives $V-m_{pg}\sim$  -0.9 mag assuming (B-V)=1.1 \citep{1961ApJ...133..869A,1995AJ....110.2885P}.
This delta is consistent with difference in the maximum light brightness of V532 Oph as seen in Figure 2,  V=11.7 mag and m$_{pg}$=12.6-12.7 mag.
Therefore, there is good agreement between the datasets from ASAS-3 and \citet{1942AnHar.109....1S}.  There is some indication that V532 Oph brightened by $\sim$0.1 mag between 1900 and 1930. The ASAS-3 lightcurve  shows evidence for pulsations with a period, $\sim$50 d when the star is at maximum light, but there aren't enough data to determine a period. This period is typical for RCB stars \citep{1990MNRAS.247...91L}. 

There is also one epoch of IJK photometry taken with Denis \citep{2005yCat.2263....0T} (JD 2451094.682577)
I = 16.561$\pm$0.10 mag, J = 13.448$\pm$0.10 mag, K = 9.333$\pm$0.07 mag, and one epoch of JHK photometry taken with 2MASS \citep{2003tmc..book.....C}
(JD 2450963.6926) J = 9.050$\pm$0.023 mag, H = 	8.784$\pm$0.051 mag, K = 8.555$\pm$0.021 mag. 
 In various other catalogs, USNO-B1.0 (Epoch 1971.5), USNO-A2.0 (Epoch 1980.858), HST GSC 1.2 (Epoch 1987.634), and HST GSC 2.2 (Epoch 1997.244), V532 Oph appears to be at or near maximum light. 
 V532 Oph was detected with IRAS \citep{1988iras....7.....H} only at 12\micron, (F = 0.54$\pm$ 0.06 Jy). Upper limits were obtained in the other bands. The IRAS flux in RCB stars is typically highest in the 12\micron~band. The RCB dust shells have temperatures in the 600-900 K range and so their emission peaks are to the blue of 12\micron~\citep{1985A&A...152...58W}. 

V532 Oph was in a deep decline ($\Delta$J$\gtrsim$4.4 mag) when the Denis data were obtained (JD 2451094; 1998 October 07). There is one major decline (JD 2425362; 1928 April 25) in the  \citet{1942AnHar.109....1S} data shown in Figure 2a and at least three major declines in the ASAS-3 data, shown in Figure 2b,  all of which go below the faint limit ($\sim$14 mag) of the telescope (ending near JD 2452600, 2002 November; JD 2453300-2453800, 2004 October - 2006 March; JD 2454400-2454600, 2007 October - 2008 May). Nearly all known RCB stars spend a majority of their 
time at maximum light with a characteristic time between irregularly spaced declines of about 1100 days \citep{1986ASSL..128..151F}. However, there is a wide variation in decline activity from star to star \citep{1996AcA....46..325J}. Also, individual stars often vary from extremely active to extremely inactive on the timescale of decades
\citep[e.g.,][]{1991AAVSM...4.....M}. V532 Oph seems to have been in an active phase over the last 10 years and was relatively inactive early in the 20th century. However, it should be pointed out that the points plotted in Figure 2a are quite widely spaced and some declines may have been missed. In a typical year, 3 or 4 plates were obtained over a 4 month period and then nothing for 8 months. 

A typical RCB star spectrum is characterized by weak or absent hydrogen lines and molecular bands (CH); strong carbon lines and molecular bands (CN, C$_2$); and little or no $^{13}$C \citep{1996PASP..108..225C, 2001ApJ...554..298A}. The continuum seen in the spectrum of V532 Oph in Figure 3 is quite red, but its molecular features are quite weak, indicating that it is a significantly reddened, warm RCB star. 
The RCB stars show a range of effective temperatures, from the warm (6000-7000 K) RCB stars which show only weak molecular bands to the cool ($\sim$5000 K) RCB stars which have much stronger molecular bands. 
The spectrum of V532 Oph is plotted along with two other warm RCB stars, W Men and HV 12842 \citep{1972MNRAS.158P..11F,1990MNRAS.245..119G}. 
The characteristic Swan bands of C$_2$ are weak but present. CN bands are probably present as well. H$\beta$, if present, is extremely weak, and there is no sign of H$\gamma$ or H$\delta$. The CH band is also extremely weak. There is a large variation in the hydrogen deficiency of the RCB stars \citep{2000A&A...353..287A}. V854 Cen, for example, which is less hydrogen deficient, clearly shows the Balmer series  \citep{1989MNRAS.238P...1K}. Also, there is no sign of the isotopic $^{12}$C$^{13}$C Swan bandhead at 4744 \AA\  in the spectrum of V532 Oph \citep{1991Obs...111..244L}. 

V532 lies at low Galactic latitude $\sim$6\arcdeg~and longitude $\sim$4\arcdeg, so there is significant reddening along the line of sight. Using the reddening maps of \citet{1998ApJ...500..525S}, the foreground extinction toward V532 Oph is estimated to be A$_V\sim$3 mag. There are only the two observations of B-V, mentioned above,  that exist for V532 Oph. They were obtained when the star was 1.1 mag below maximum as it recovered from a decline. The measured value of (B-V) = 1.1 will be the intrinsic stellar color reddened by a combination of circumstellar and interstellar dust. If V532 is a warm RCB star, as indicated by its spectrum, (T$_{eff}\sim$ 7000 K), then (B-V)$_o\sim$0.5 \citep{1990MNRAS.247...91L}. Then E(B-V)= 0.6 which implies A$_V\sim$2 mag (assuming R$_V$=3.1). 
The absolute magnitude of V532 Oph should be M$_V\sim$-5 mag  \citep{1996PASP..108..225C, 2001ApJ...554..298A}, so assuming V$_{max}$=11.7 and a reddening of A$_V$=2-3 mag, then it lies at a distance of 5.5-8.7 kpc. Therefore, this star may lie in the Galactic bulge similar to some other recently discovered RCB stars \citep{2005AJ....130.2293Z,2008A&A...481..673T}. 

The new data presented here clearly show that V532 Oph has all the characteristics of a typical RCB star including sudden deep irregularly spaced declines, an optical spectrum showing weak or absent hydrogen lines, the Swan bands of C$_2$, and no evidence for $^{13}$C. In addition, the star shows small pulsations typical of an RCB star and an infrared excess due to circumstellar dust. The combination of long-term photometric coverage, visible spectra and IR photometry, used here for the definitive identification of V532 Oph as an RCB star, can be easily applied to other RCB candidates in the future now that all-sky survey telescopes such as ASAS-3 are available. 



\acknowledgments
This paper is partly based on work supported financially by the National Research Foundation of South Africa. The research
of DLW is supported by a Discovery Grant from NSERC - the Natural Sciences and Engineering Research Council of Canada.

\bibliography{/Users/gclayton/projects/latexstuff/everything2}

\begin{thebibliography}{35}
\expandafter\ifx\csname natexlab\endcsname\relax\def\natexlab#1{#1}\fi
\expandafter\ifx\csname href\endcsname\relax
  \def\href#1#2{}\fi
\expandafter\ifx\csname urllinklabel\endcsname\relax
  \def\urllinklabel{[LINK]}\fi
\expandafter\ifx\csname adsurllinklabel\endcsname\relax
  \def\adsurllinklabel{[ADS]}\fi

\bibitem[{{Alcock et al.}(2001)}]{2001ApJ...554..298A}
{Alcock et al.} 2001, \apj, 554, 298


\bibitem[{{Arp}(1961)}]{1961ApJ...133..869A}
{Arp}, H. 1961, \apj, 133, 869


\bibitem[{{Asplund} {et~al.}(2000){Asplund}, {Gustafsson}, {Lambert}, \&
  {Rao}}]{2000A&A...353..287A}
{Asplund}, M., {Gustafsson}, B., {Lambert}, D.~L., \& {Rao}, N.~K. 2000, \aap,
  353, 287


\bibitem[{{Benson} {et~al.}(1994){Benson}, {Clayton}, {Garnavich}, \&
  {Szkody}}]{1994AJ....108..247B}
{Benson}, P.~J., {Clayton}, G.~C., {Garnavich}, P., \& {Szkody}, P. 1994, \aj,
  108, 247


\bibitem[{{Clayton}(1996)}]{1996PASP..108..225C}
{Clayton}, G.~C. 1996, \pasp, 108, 225


\bibitem[{{Clayton} {et~al.}(2007){Clayton}, {Geballe}, {Herwig}, {Fryer}, \&
  {Asplund}}]{2007ApJ...662.1220C}
{Clayton}, G.~C., {Geballe}, T.~R., {Herwig}, F., {Fryer}, C., \& {Asplund}, M.
  2007, \apj, 662, 1220


\bibitem[{{Clayton} {et~al.}(2002){Clayton}, {Hammond}, {Lawless}, {Kilkenny},
  {Evans}, {Mattei}, \& {Landolt}}]{2002PASP..114..846C}
{Clayton}, G.~C., {Hammond}, D., {Lawless}, J., {Kilkenny}, D., {Evans}, T.~L.,
  {Mattei}, J., \& {Landolt}, A.~U. 2002, \pasp, 114, 846


\bibitem[{{Clayton} {et~al.}(2005){Clayton}, {Herwig}, {Geballe}, {Asplund},
  {Tenenbaum}, {Engelbracht}, \& {Gordon}}]{2005ApJ...623L.141C}
{Clayton}, G.~C., {Herwig}, F., {Geballe}, T.~R., {Asplund}, M., {Tenenbaum},
  E.~D., {Engelbracht}, C.~W., \& {Gordon}, K.~D. 2005, \apjl, 623, L141


\bibitem[{{Cutri et al.}(2003)}]{2003tmc..book.....C}
{Cutri et al.} 2003, {2MASS All Sky Catalog of point sources.} (The IRSA 2MASS
  All-Sky Point Source Catalog, NASA/IPAC Infrared Science
  Archive.~http://irsa.ipac.caltech.edu/applications/Gator/)


\bibitem[{{Feast}(1972)}]{1972MNRAS.158P..11F}
{Feast}, M.~W. 1972, \mnras, 158, 11P


\bibitem[{{Feast}(1986)}]{1986ASSL..128..151F}
---. 1986, IAU Colloq. 87, 128, p. 151


\bibitem[{{Fujimoto}(1977)}]{1977PASJ...29..331F}
{Fujimoto}, M.~Y. 1977, \pasj, 29, 331


\bibitem[{{Goldsmith} {et~al.}(1990){Goldsmith}, {Evans}, {Albinson}, \&
  {Bode}}]{1990MNRAS.245..119G}
{Goldsmith}, M.~J., {Evans}, A., {Albinson}, J.~S., \& {Bode}, M.~F. 1990,
  \mnras, 245, 119


\bibitem[{{Helou} \& {Walker}(1988)}]{1988iras....7.....H}
{Helou}, G. \& {Walker}, D.~W., eds. 1988, {Infrared astronomical satellite
  (IRAS) catalogs and atlases. Volume 7: The small scale structure catalog},
  Vol.~7


\bibitem[{{Hesselbach} {et~al.}(2003){Hesselbach}, {Clayton}, \&
  {Smith}}]{2003PASP..115.1301H}
{Hesselbach}, E., {Clayton}, G.~C., \& {Smith}, P.~S. 2003, \pasp, 115, 1301


\bibitem[{{Iben} {et~al.}(1996){Iben}, {Tutukov}, \&
  {Yungelson}}]{1996ApJ...456..750I}
{Iben}, I.~J., {Tutukov}, A.~V., \& {Yungelson}, L.~R. 1996, \apj, 456, 750


\bibitem[{{Jurcsik}(1996)}]{1996AcA....46..325J}
{Jurcsik}, J. 1996, Acta Astronomica, 46, 325


\bibitem[{{Kilkenny} \& {Marang}(1989)}]{1989MNRAS.238P...1K}
{Kilkenny}, D. \& {Marang}, F. 1989, \mnras, 238, 1P


\bibitem[{{Lawson} {et~al.}(1990){Lawson}, {Cottrell}, {Kilmartin}, \&
  {Gilmore}}]{1990MNRAS.247...91L}
{Lawson}, W.~A., {Cottrell}, P.~L., {Kilmartin}, P.~M., \& {Gilmore}, A.~C.
  1990, \mnras, 247, 91


\bibitem[{{Lloyd Evans} {et~al.}(1991){Lloyd Evans}, {Kilkenny}, \& {van
  Wyk}}]{1991Obs...111..244L}
{Lloyd Evans}, T., {Kilkenny}, D., \& {van Wyk}, F. 1991, The Observatory, 111,
  244


\bibitem[{{Mattei} {et~al.}(1991){Mattei}, {Waagen}, \&
  {Foster}}]{1991AAVSM...4.....M}
{Mattei}, J.~A., {Waagen}, E.~O., \& {Foster}, E.~G. 1991, AAVSO Monogr.,
  No.~4,, 4


\bibitem[{{Pierce} \& {Jacoby}(1995)}]{1995AJ....110.2885P}
{Pierce}, M.~J. \& {Jacoby}, G.~H. 1995, \aj, 110, 2885


\bibitem[{{Pojmanski}(2002)}]{2002AcA....52..397P}
{Pojmanski}, G. 2002, Acta Astronomica, 52, 397


\bibitem[{{Renzini}(1979)}]{1979sss..meet..155R}
{Renzini}, A. 1979, in ASSL Vol. 75: Stars and star systems, ed. B.~E.
  {Westerlund}, 155--171


\bibitem[{{Saio} \& {Jeffery}(2002)}]{2002MNRAS.333..121S}
{Saio}, H. \& {Jeffery}, C.~S. 2002, \mnras, 333, 121


\bibitem[{{Samus} {et~al.}(2009){Samus}, {Durlevich}, \& {et
  al.}}]{2009yCat....102025S}
{Samus}, N.~N., {Durlevich}, O.~V., \& {et al.} 2009, VizieR Online Data
  Catalog, 1, 2025


\bibitem[{{Schlegel} {et~al.}(1998){Schlegel}, {Finkbeiner}, \&
  {Davis}}]{1998ApJ...500..525S}
{Schlegel}, D.~J., {Finkbeiner}, D.~P., \& {Davis}, M. 1998, \apj, 500, 525


\bibitem[{{Swope}(1942)}]{1942AnHar.109....1S}
{Swope}, H.~H. 1942, Annals of Harvard College Observatory, 109, 1


\bibitem[{{The Denis Consortium}(2005)}]{2005yCat.2263....0T}
{The Denis Consortium}. 2005, VizieR Online Data Catalog, 2263, 0


\bibitem[{{Tisserand} {et~al.}(2004){Tisserand}, {Marquette}, {Beaulieu}, {de
  Laverny}, {Lesquoy}, {Milsztajn}, {Afonso}, {Albert}, {Andersen}, {Ansari},
  {Aubourg}, {Bareyre}, {Bauer}, {Blanc}, {Charlot}, {Coutures}, {Derue},
  {Ferlet}, {Fouqu{\'e}}, {Glicenstein}, {Goldman}, {Gould}, {Graff}, {Gros},
  {Haissinski}, {Hamadache}, {de Kat}, {Lasserre}, {Le Guillou}, {Loup},
  {Magneville}, {Mansoux}, {Maurice}, {Maury}, {Moniez},
  {Palanque-Delabrouille}, {Perdereau}, {Pr{\'e}vot}, {Rahal}, {Regnault},
  {Rich}, {Spiro}, {Vidal-Madjar}, {Vigroux}, \&
  {Zylberajch}}]{2004A&A...424..245T}
{Tisserand}, P., {Marquette}, J.~B., {Beaulieu}, J.~P., {de Laverny}, P.,
  {Lesquoy}, {\'E}., {Milsztajn}, A., {Afonso}, C., {Albert}, J.~N.,
  {Andersen}, J., {Ansari}, R., {Aubourg}, {\'E}., {Bareyre}, P., {Bauer}, F.,
  {Blanc}, G., {Charlot}, X., {Coutures}, C., {Derue}, F., {Ferlet}, R.,
  {Fouqu{\'e}}, P., {Glicenstein}, J.~F., {Goldman}, B., {Gould}, A., {Graff},
  D., {Gros}, M., {Haissinski}, J., {Hamadache}, C., {de Kat}, J., {Lasserre},
  T., {Le Guillou}, L., {Loup}, C., {Magneville}, C., {Mansoux}, B., {Maurice},
  {\'E}., {Maury}, A., {Moniez}, M., {Palanque-Delabrouille}, N., {Perdereau},
  O., {Pr{\'e}vot}, L., {Rahal}, Y., {Regnault}, N., {Rich}, J., {Spiro}, M.,
  {Vidal-Madjar}, A., {Vigroux}, L., \& {Zylberajch}, S. 2004, \aap, 424, 245


\bibitem[{{Tisserand et al.}(2008)}]{2008A&A...481..673T}
{Tisserand et al.} 2008, \aap, 481, 673


\bibitem[{{Walker}(1985)}]{1985A&A...152...58W}
{Walker}, H.~J. 1985, \aap, 152, 58


\bibitem[{{Webbink}(1984)}]{1984ApJ...277..355W}
{Webbink}, R.~F. 1984, \apj, 277, 355


\bibitem[{{Zacharias} {et~al.}(2004){Zacharias}, {Urban}, {Zacharias},
  {Wycoff}, {Hall}, {Monet}, \& {Rafferty}}]{2004AJ....127.3043Z}
{Zacharias}, N., {Urban}, S.~E., {Zacharias}, M.~I., {Wycoff}, G.~L., {Hall},
  D.~M., {Monet}, D.~G., \& {Rafferty}, T.~J. 2004, \aj, 127, 3043


\bibitem[{{Zaniewski} {et~al.}(2005){Zaniewski}, {Clayton}, {Welch}, {Gordon},
  {Minniti}, \& {Cook}}]{2005AJ....130.2293Z}
{Zaniewski}, A., {Clayton}, G.~C., {Welch}, D.~L., {Gordon}, K.~D., {Minniti},
  D., \& {Cook}, K.~H. 2005, \aj, 130, 2293


\end{thebibliography}

\clearpage

\begin{figure}
\figurenum{1} 
\begin{center}
\includegraphics[width=5in,angle=0]{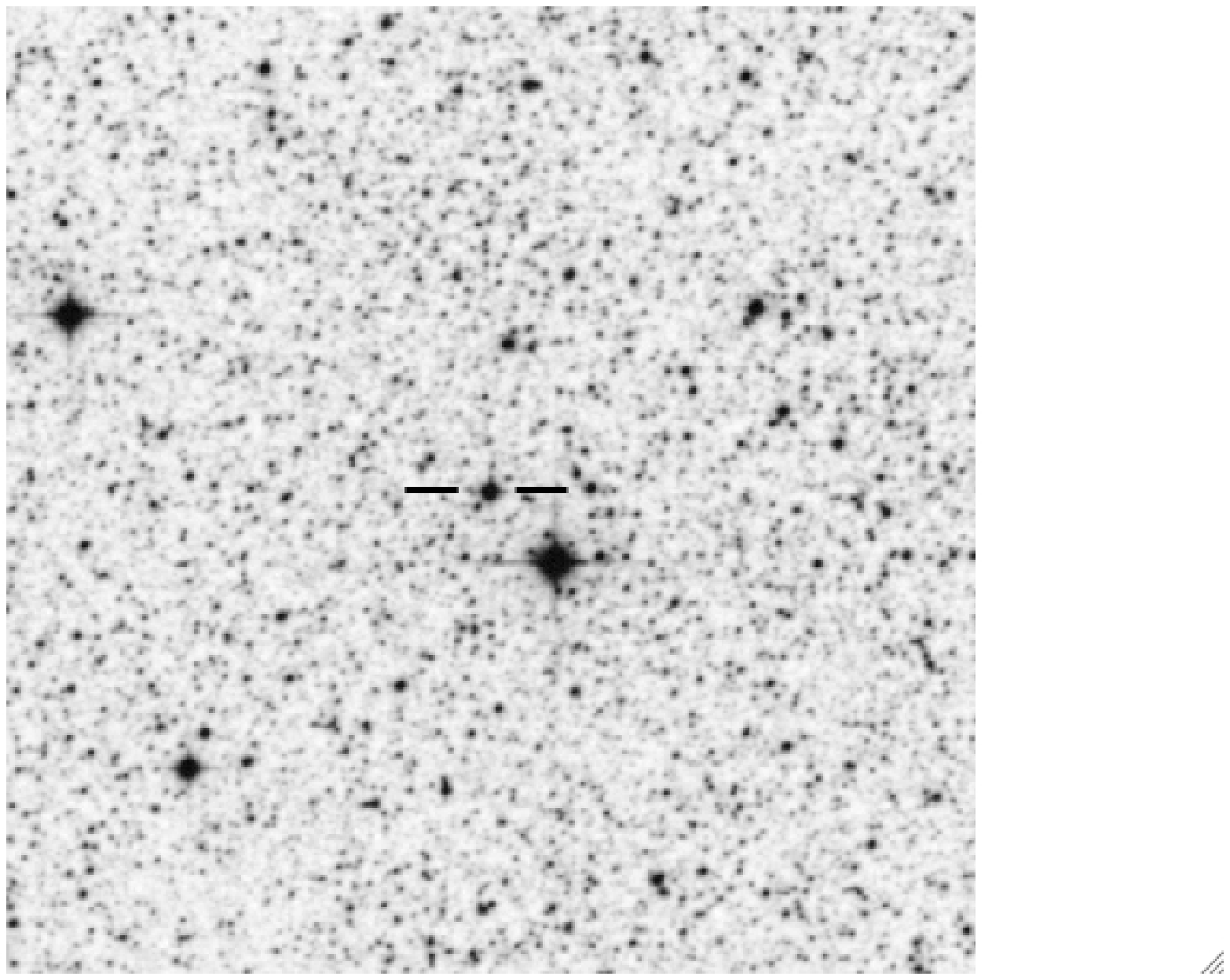}
\end{center}
\caption{UKSTU Schmidt Red Sky Survey plate (XS588 A1BQ, epoch 1992.55) with V532 Oph marked. The chart is 10\arcmin  $\times$ 10\arcmin. North is up and east is left.}
\end{figure}

\begin{figure}
\figurenum{2a} 
\includegraphics[width=7in,angle=0]{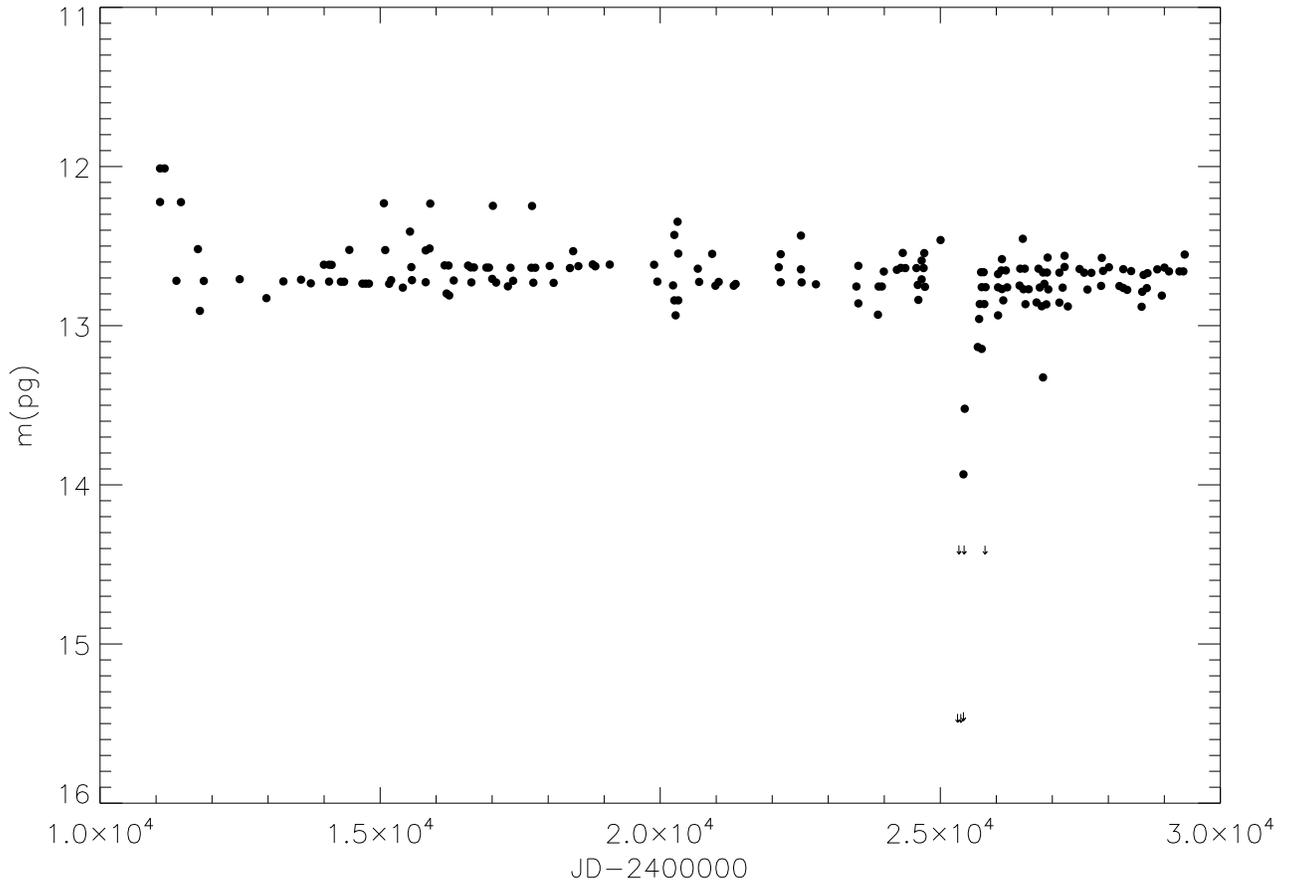}
\caption{Lightcurve of V532 Oph from 1890 to 1939. The filled circles are photometry from Harvard College Observatory plates  \citep{1942AnHar.109....1S}. The arrows are upper limits.}
\end{figure}

\begin{figure}
\figurenum{2b} 
\includegraphics[width=7in,angle=0]{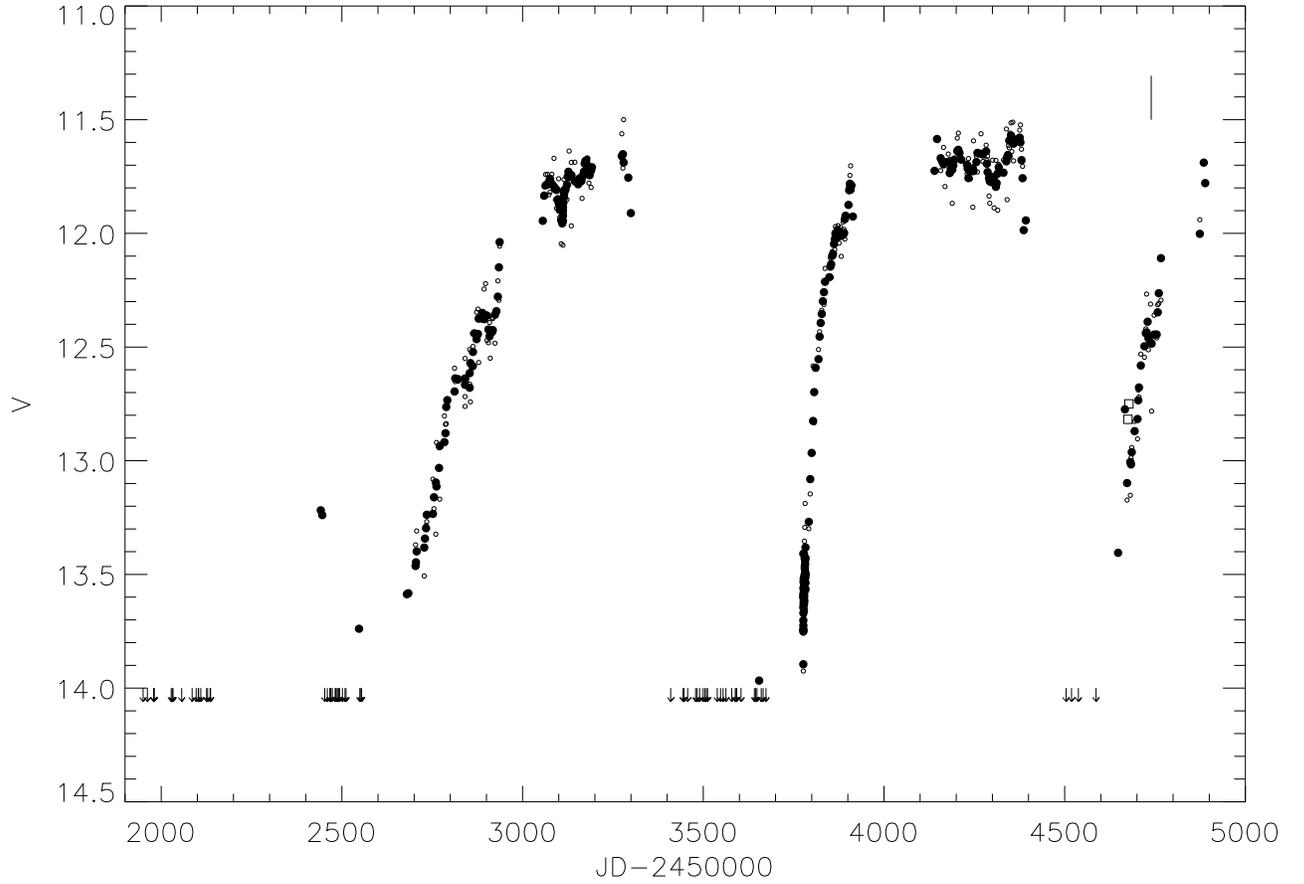}
\caption{Lightcurve of V532 Oph from 2001 to 2009. The filled circles are the ASAS-3 V-band data with a 5-point smooth. The small open circles represent the unsmoothed data. The arrows are upper limits. The open squares are AAVSO V-band data. The vertical bar marks the epoch in which the spectra were obtained.}
\end{figure}

\begin{figure}
\figurenum{3} 
\includegraphics[width=7in,angle=0]{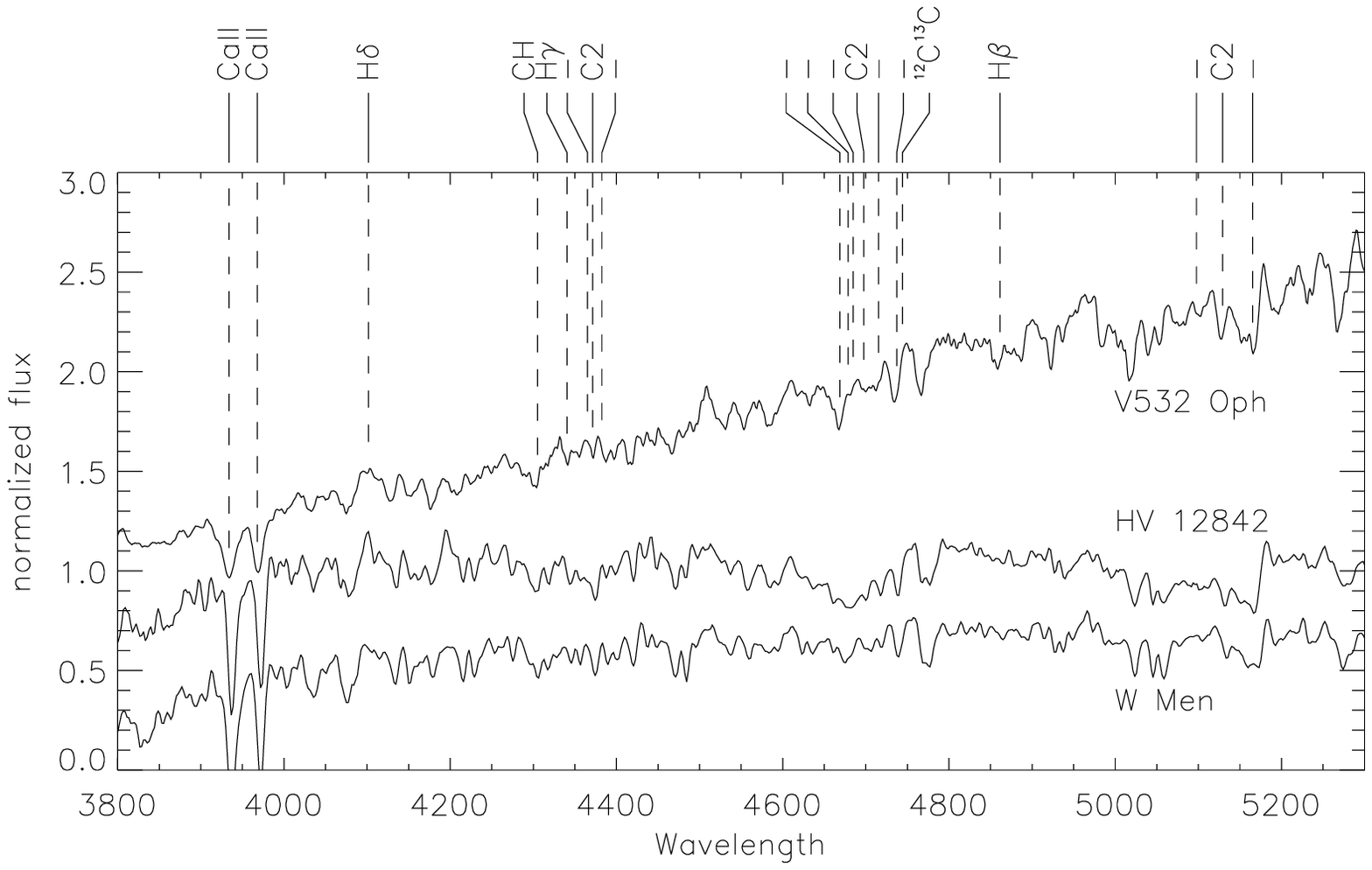}
\caption{Spectrum of V532 Oph obtained in 2008 September produced from the median of eight individual spectra. Also shown are the spectra of the RCB stars, W Men and HV 12842 for comparison. The spectra have been normalized and shifted for easy comparison. }
\end{figure}


\end{document}